\RequirePackage{fix-cm}
\documentclass[smallextended]{svjour3}       
\smartqed  
\usepackage{graphicx}
\usepackage{amssymb} 

\usepackage{subfigure}
\usepackage{amsmath}
\usepackage{amsfonts}
\usepackage{amsbsy}
\usepackage{times}
\usepackage{mathptmx}
\usepackage[left]{lineno}
\usepackage{appendix}
\usepackage{color} 
 \journalname{J. Braz. Soc. Mech. Sci. Eng.}
\begin{document}

\title{Numerical solution for Kapitza waves on a thin liquid film\thanks{This is a pre-print of an article published in Journal of the Brazilian Society of Mechanical Sciences and Engineering, 40:375, 2018. The final authenticated version is available online at: http://dx.doi.org/10.1007/s40430-018-1295-1}
}


\author{Bruno Pelisson Chimetta \and
        Mohammad Zakir Hossain \and
        Erick de Moraes Franklin 
}


\institute{Bruno Pelisson Chimetta \at
              School of Mechanical Engineering, University of Campinas - UNICAMP, Brazil \\
							 \email{brunopchimetta@gmail.com}               
									 \and
					 Mohammad Zakir Hossain \at
              Department of Mechanical and Materials Engineering, The University of Western Ontario, Canada \\
							 \email{mhossa7@uwo.ca}														
									 \and
           Erick de Moraes Franklin (corresponding author)\at
              School of Mechanical Engineering, University of Campinas - UNICAMP, Brazil \\
              Tel.: +55-19-35213375\\							
              orcid.org/0000-0003-2754-596X\\
              \email{franklin@fem.unicamp.br}             
}

\date{Received: date / Accepted: date}

\maketitle

\begin{abstract}
\begin{sloppypar}
The flow of a liquid film over an inclined plane is frequently found in nature and industry, and, under some conditions, instabilities in the free surface may appear. These instabilities are initially two-dimensional surface waves, known as Kapitza waves. Surface waves are important to many industrial applications. For example, liquid films with surface waves are employed to remove heat from solid surfaces. The initial phase of the instability is governed by the Orr-Sommerfeld equation and the appropriate boundary conditions; therefore, the fast and accurate solution of this equation is useful for industry. This paper presents a spectral method to solve the Orr-Sommerfeld equation with free surface boundary conditions. Our numerical approach is based on a Galerkin method with Chebyshev polynomials of the first kind, making it possible to express the Orr-Sommerfeld equation and their boundary conditions as a generalized eigenvalue problem. The main advantages of the present spectral method when compared to others such as, for instance, spectral collocation, are its stability and its readiness in including the boundary conditions in the discretized equations. We compare our numerical results with analytical solutions based on Perturbation Methods, which are valid only for long wave instabilities, and show that the results agree in the region of validity of the long-wave hypothesis. Far from this region, our results are still valid. In addition, we compare our results with published experimental results and the agreement is very good. The method is stable, fast, and capable to solve initial instabilities in free surface flows.
\end{sloppypar}
\end{abstract}

\section{Introduction}

The flow of a liquid film over inclined and vertical surfaces is frequently found in nature and industry. Water flows over non permeable grounds, resin and ink flows over painted plates, and water flows over heated domes are some examples of the flow of liquid films. Under some conditions, instabilities that are initially two-dimensional surface waves, known as Kapitza waves, may appear and evolve to three-dimensional forms. The presence of surface waves is important to many industrial applications. For instance, it enhances heat transfers, and, therefore, it is sought for some applications using liquid films to remove heat from solid surfaces. On the other hand, if the application involves painting or surface coating, surface waves are not desired. 

The initial phase of the instability is governed by the Orr-Sommerfeld equation and the appropriate boundary conditions; therefore, the fast and accurate solution of this equation is useful for industry. A fast solution may be incorporated to controlling systems used in, for example, painting and coating devices, and refrigeration devices of nuclear facilities, in order to change in a few seconds the inclination, the flow rate or any other relevant parameter that must be changed to allow or avoid the growth of surface waves. 

By the end of the first half of the twentieth century, the study on the flow of liquid films increased in importance due to new industrial processes emerging in that period. One of the first researchers to investigate this problem was Kapitza \cite{kapitsa1948wave,kapitza1949wave}, who performed experimental and theoretical works on the flow of liquid films on a vertical wall. He proposed that the ratio between the effects of surface tension and inertia is the pertinent dimensionless number, known as Kapitza number, which is an indicator of the hydrodynamic regime of the flow. Some years later, Benjamin \cite{benjamin1957wave} presented a theoretical work on the flow of liquid films over an inclined plane at low Reynolds numbers. He wrote the wavenumber as a function of the Reynolds number, and obtained the neutral stability curves for flows on a vertical wall. In the same period, Yih \cite{yih1963stability}, by defining the perturbation by means of a stream function, obtained long and short wave solutions for the flow of liquid films at low Reynolds number.
 
Benney \cite{benney1966long} presented an asymptotic analysis for long waves appearing on liquid films. By performing a third order power expansion of the perturbed free surface, he solved an eigenvalue problem and obtained analytical expressions for the growth rate, wavelength, and celerity of long wave instabilities. In addition, from his solutions it is possible to find the Reynolds numbers from which the film is linearly unstable for long waves.

The first numerical solution of the Orr-Sommerfeld equation was obtained by Thomas \cite{thomas1953stability} in an attempt to solve the controversies that existed in that period about the validity of the asymptotic methods. Thomas used the finite difference method, replacing the fourth-order differential equation with a system of differences of the same order, but with a truncation error involving an eighth-order derivative. To solve the linear algebraic system, Thomas used the Gaussian elimination method, and obtained a critical Reynolds number equal to $5780$, confirming that the plane Poiseuille flow was in fact unstable, in agreement with Lin \cite{lin1946stability}. A few years later, Dolph and Lewis \cite{dolph1958application} solved the Poiseuille stability problem using a numerical method based on an expansion of orthogonal functions. For a characteristic wavenumber equal to $1$ and a number of terms $N$ in the expansion equal to 20, they found a critical Reynolds number equal to $5800$, a result which, despite the small discrepancy, agrees with \cite{thomas1953stability}.

Orzag \cite{orszag1971accurate} used Chebyshev polynomials together with the implementation of a QR algorithm to solve numerically the Orr-Sommerfeld equation applied to the stability of plane Poiseuille flows. With this approach, great accuracy is achieved with smaller computational time when compared to previous methods. He found a critical Reynolds number of $5772.22$ for a wavenumber equal to $1.02056$. The method developed by Orzag \cite{orszag1971accurate} proved to be one of the best methods for solving stability problems due to its high accuracy and the low computational cost, being frequently used until today.

Many of the numerical approaches employed until the early 1980's had, as a common feature, the use of shooting methods. Despite their simple implementation, they lead to convergence problems when the initial guess is far from the solution, as well as when the Reynolds number assumes high values.

\begin{sloppypar}
Floryan et al. \cite{floryan1987instabilities} used a Newton-Raphson method with ortho-normalization to investigate the stability of a liquid film on an inclined plane. They obtained the critical Reynolds numbers for different values of surface tension, and found that the growth rate decreases as the surface tension increases or the plane angle decreases. They found also that, at the limit of high values of the Reynolds numbers, the shear mode of the liquid film is inviscidly stable regardless of the magnitude of the surface tension.
\end{sloppypar}

Liu et al. \cite{liu1993measurements} investigated experimentally the primary instabilities of thin liquid films flowing over an inclined plane. The authors used water and glycerin-water solutions to vary the properties of the liquid, they forced small pressure variations at the film inlet, and they measured the surface waves with light emitters and receptors. Liu et al. \cite{liu1993measurements} found the critical Reynolds number as a function of the slope angle for the onset of the waves, and also the growth rate and wave velocities as functions of the wavenumber. They found good agreement with the linear theory, specially for the critical Reynolds number. 

Kalliadasis et al. \cite{kalliadasis2003thermocapillary} studied the flow of a liquid film over an inclined plane uniformly heated at moderate Reynolds numbers (between 10 and 30). The computations were conducted by using an IBL (Integral Boundary Layer) approximation of the Navier-Stokes and energy equations together with free surface boundary conditions. The authors analyzed the linear stability with respect to both two-dimensional and three-dimensional perturbations, and found that an increase in the slope or in the Marangoni number increases the unstable region, where the Marangoni number expresses the relative importance of thermocapilarity and viscous tensions. In addition, Kalliadasis et al. \cite{kalliadasis2003thermocapillary} found that inertia dominates Marangoni forces for large film thicknesses and small interfacial deformations, while Marangoni forces dominate inertia for small thicknesses and large deformations.

\begin{sloppypar}
Wierschem and Askel \cite{wierschem2003instability} investigated numerically the stability of a liquid film flowing over an undulated surface. The geometry of the inclined surface was $A sin(2 \pi x / \lambda)$, where $\lambda$ is the wavelength and $A$ the amplitude of the undulation, and $x$ is the coordinate in the main direction of the flow.  To find the unstable modes, the authors found the steady state solution, expanded the perturbations in power series of the characteristic wavenumber, and performed a spatial analysis. Wierschem and Askel \cite{wierschem2003instability} found that the critical Reynolds number of long waves for the flow over a corrugated wall is larger than that over a flat wall.
\end{sloppypar}

Baxter et al. \cite{baxter2009three} investigated the Stokes flow of a liquid film over an inclined plane in the presence of a fixed obstacle. The authors presented the boundary conditions of the flow over and around the obstacle, followed by an asymptotic analysis for the case of the flow over the obstacle. Next, they used two numerical approaches to analyze the problem, the first an approximation by finite differences and the second an interpolation by Hermitian functions. Baxter et al. \cite{baxter2009three} used both approaches to evaluate the curvature of the interface in the presence of the obstacle, and showed that a flow over and around a truncated cylinder indicates the possibility of two solutions. This result imply that, for this case, the steady state flow is dependent on the initial conditions.

Liu and Liu \cite{liu2009instabilities} studied the flow of liquid films over an inclined plane with a porous surface. The inclined plane consisted of an impermeable wall with a homogeneous porous layer over it; therefore, the liquid film flowed over and within the porous layer. The authors used the Navier-Stokes equations for the liquid film above the porous layer, and the Darcy equation for the flow within the porous layer. To solve the perturbed equations, Liu and Liu \cite{liu2009instabilities} expanded the amplitudes of normal modes in series of Chebyshev polynomials, so that the equations led to a generalized sixth order eigenvalue problem, which was solved using a collocation method. The authors used 60 polynomials in their numerical computations. Liu and Liu \cite{liu2009instabilities} also concluded that an exchange between the porous and film layers occurs as the Reynolds number increases, with permeability being one of the main factors to determine the instability of some modes.

Recently, Rohlfs et al. \cite{rohlfs2017hydrodynamic} investigated the flow of a liquid film down the underside of an inclined plane. The authors used both the weighted-integral boundary layer model and direct numerical simulation -- DNS. They studied the effect of the Rayleigh-Taylor instability that lead to the formation of two- and three-dimensional waves, and possible dripping, at the surface of the liquid layer. The DNS approach was implemented together with a scheme of interface compression in order to reduce the amplitude of artificial velocities at the interface, and allow a precise prediction of falling liquid films with high wave amplitudes. The authors found good agreement between the model and DNS under destabilizing gravity conditions, including the wave celerity and the wave peak height. Rohlfs et al. \cite{rohlfs2017hydrodynamic} showed that the increase in the Reynolds number implies a decrease of the maximum film thickness.

This paper addresses the initial instabilities of a liquid film flowing over an inclined plane, and presents a spectral method to solve the Orr-Sommerfeld equation with free surface boundary conditions. Our numerical approach is based on a Galerkin method with Chebyshev polynomials of the first kind, making it possible to express the Orr-Sommerfeld equation and their boundary conditions as a generalized eigenvalue problem. Different from other spectral methods, the present one is at the same time stable and straightforward in including the boundary conditions in the discretized equations. We compare our numerical results with analytical solutions based on Perturbation Methods, which are valid only for long wave instabilities, and show that the results agree in the region of validity of the long-wave hypothesis. Far from this region, our results are still valid. In addition, we compare our results with the experimental results of Liu et al. \cite{liu1993measurements} and show that the agreement between them is very good. The method is stable, fast, and capable to solve initial instabilities in free surface flows.

Section \ref{section:linear_stability} describes the physics and main equations involved in the linear stability analysis, Section \ref{section:numerical} presents the numerical method used in this study, Section \ref{section:results} presents the results of the stability analysis and compares them with a long-wave analytical solution and with the experimental results of Liu et al. \cite{liu1993measurements}, and Section \ref{section:conclusions} concludes the paper. The Appendix \ref{appendix_A} presents the long-wave solution obtained by a Perturbation Method.

\section{Linear stability equations}
\label{section:linear_stability}

\begin{figure}[ht]	
 	\centering
 	\includegraphics[width=0.85\columnwidth]{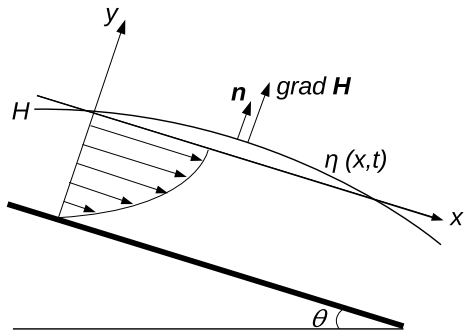}
 	\caption{Layout of the falling film.}
 	\label{fig:1}
\end{figure}

We consider a liquid film of thickness $h$ falling down on an inclined plane with an angle $\theta$ with respect to the horizontal. The flow is driven by gravity and the only opposing force is the friction between the fluid and the inclined plane. The free surface is initially flat and the interface between the liquid and the gas has a surface tension $\gamma$. The gas pressure is uniform and equal to $P_{0}$. The fluid is considered Newtonian, with viscosity $\mu$ and density $\rho$. Fig. \ref{fig:1} presents a layout of the considered problem.

We analyze next the stability of the free surface with regard to the formation of Kapitza waves. The equations for the linear stability analysis are presented in Subsection \ref{section:equations} and the perturbed equations in Subsection \ref{section:perturbations}. The equations are in a two-dimensional space, which is justified by Squire's theorem \cite{Drazin_Reid}.

\subsection{Conservation equations}
\label{section:equations}

The two-dimensional mass and momentum conservation equations applied to the liquid film are given by Eqs. \ref{mass}, \ref{momentum_x} and \ref{momentum_y} \cite{Batchelor,Livre_Charru}

\begin{equation}
	\frac{\partial u}{\partial x} + \frac{\partial v}{\partial y} = 0
	\label{mass}
\end{equation}

\begin{equation}
	\rho \Biggl[\frac{\partial u}{\partial t} + u\frac{\partial u}{\partial x} + v\frac{\partial u}{\partial y}\Biggl] = - \frac{\partial p}{\partial x} + \mu \Biggl(\frac{\partial ^{2} u}{\partial x^{2}} + \frac{\partial ^{2} u}{\partial y^{2}}\Biggl) + \rho g_{x}
	\label{momentum_x}
\end{equation}

\begin{equation}
	\rho \Biggl[\frac{\partial v}{\partial t} + u\frac{\partial v}{\partial x} + v\frac{\partial v}{\partial y}\Biggl] = - \frac{\partial p}{\partial y} + \mu \Biggl(\frac{\partial ^{2} v}{\partial x^{2}} + \frac{\partial ^{2} v}{\partial y^{2}}\Biggl) + \rho g_{y}
	\label{momentum_y}
\end{equation}

\noindent where $x$ and $y$ are the longitudinal and transverse coordinates, $u$ and $v$ are the longitudinal and transverse components of the fluid velocity, and $g_x$ and $g_y$ are the longitudinal and transverse components of the gravity acceleration, respectively, and $t$ is the time.

Based on the conservation equations, and considering the boundary conditions of (i) permanent flow just upstream of the considered domain, (ii) non-slip at solid surface, and (iii) continuous velocity, no shear and constant pressure at the liquid-gas interface, we find a solution corresponding to a steady parallel flow, with a planar interface and parabolic velocity profile \cite{Livre_Charru}, given by Eqs. \ref{U_basic} and \ref{V_basic},

\begin{equation}
	\overline{U}(y) = U_{0}(1 - \frac{y^{2}}{h^{2}})
	\label{U_basic}
\end{equation}

\begin{equation}
	\overline{V}(y) = 0
	\label{V_basic}
\end{equation}

\noindent where the fluid velocity at the interface $U_{0}$ is given by,

\begin{equation}
	U_{0} = \frac{\rho g h^{2} \sin (\theta)}{2 \mu}
	\label{U0_basic}	
\end{equation}

\noindent and the pressure distribution is given by Eq. \ref{P_basic},

\begin{equation}
	\overline{P}(y) = P_{0} - \rho g \cos (\theta) y
	\label{P_basic}
\end{equation}

The contributions of inertia related to viscosity, gravity, and surface tension are given, respectively, by the Reynolds $Re$, Froude $Fr$, and Weber $We$ numbers, which are dimensionless groups defined in Eq. \ref{dimensionless}
	
\begin{equation}
	Re = \frac{\rho U_{0} h}{\mu},\   \ Fr = \frac{U_{0}^{2}}{g h \cos (\theta)} = \frac{Re\ \tan (\theta)}{2},\   \ We = \frac{\rho U_{0}^{2} h}{\gamma}
	\label{dimensionless}
\end{equation}

\noindent where the Froude number is defined using the gravity component normal to the flow direction, $g \cos (\theta)$. When the interface is disturbed ($\eta (x,t) \neq 0$), the velocity profile no longer has an exact parabolic behavior and surface waves may grow if the flow has sufficient inertia.

\subsection{Perturbations}
\label{section:perturbations}

Let's impose a perturbation whose longitudinal and transverse components are $\hat{u}$ and $\hat{v}$ on a stable velocity field, so that $u = \overline{U} + \hat{u}$ and $v = 0 + \hat{v}$, and consider a streamfunction $\Psi$ associated with the perturbation, given by  $\hat{u} = \partial_y \Psi$ and $\hat{v} = - \partial_x \Psi$. The expected solutions obtained by injecting those forms in Eqs. \ref{mass} to \ref{momentum_y} and linearizing them are plane waves; therefore, the streamfunction is given by the normal modes shown in Eq. \ref{streamfunction}

\begin{equation}
\Psi (x,y,t) = \hat{\Psi}(y) e^{i\alpha (x - ct)}
\label{streamfunction}
\end{equation}

\noindent where $\alpha = kh \in \mathbb{R}$, $c = \frac{\omega}{k} \in \mathbb{C}$, $k$ is the wavenumber and $\omega$ is the angular frequency. We consider $c = c_{r} + ic_{i}$, where $c_{r}$ and $\sigma = \alpha c_{i}$ are the phase velocity and the growth rate, respectively. If $c_{i} < 0$ the system is stable, and if $c_{i} > 0$ the system is linearly unstable. The resulting equation is the Orr-Sommerfeld equation, given by Eq. \ref{Orr-Sommerfeld}

\begin{equation}
(D^{2} - \alpha ^{2})^{2}\hat{\Psi}(y) = i\alpha Re[(\overline{U} - c)(D^{2} - \alpha ^{2}) - D^{2}\overline{U}]\hat{\Psi}(y)
\label{Orr-Sommerfeld}
\end{equation}

\noindent where $D = \partial_y$. The boundary conditions are non-slip at solid surface, given by Eqs. \ref{cc1} and \ref{cc2}, and continuous velocity, constant pressure and no shear stress at the liquid-gas interface, given by Eqs. \ref{cc3}, \ref{cc4} and \ref{cc5}, respectively.

\begin{equation}
D\hat{\Psi}(-1) = 0
\label{cc1}
\end{equation}

\begin{equation}
\hat{\Psi}(-1) = 0
\label{cc2}
\end{equation}

\begin{equation}
\hat{\Psi}(0) - (c - 1)\hat{\eta} = 0
\label{cc3}
\end{equation}

\begin{equation}
D^{2}\hat{\Psi}(0) + \alpha ^{2} \hat{\Psi}(0) + \hat{\eta} D^{2}\overline{U}(0) = 0
\label{cc4}
\end{equation}

\begin{equation}
- D^3 \hat{\Psi}(0) + \left[ 3 \alpha^2 - i \alpha Re (c - 1 ) \right] D \hat{\Psi}(0) + i \alpha Re \Biggl[\frac{1}{Fr} + \frac{\alpha ^{2}}{We} \Biggl] \hat{\eta} = 0
\label{cc5}
\end{equation}

We present next a numerical method to solve the Orr-Sommerfeld equation, Eq. \ref{Orr-Sommerfeld}, with the boundary conditions given by Eqs. \ref{cc1} to \ref{cc5}.

\section{Numerical Method}
\label{section:numerical}

We used a Weighted Residual method to solve the problem numerically. For this, we assumed that the problem is governed by a linear differential equation, given by Eq. \ref{sn6}, in a domain with boundary conditions given by Eq. \ref{sn7}. The approximate solution is then given by Eq. \ref{sn8},

\begin{equation}
L(u) = 0
\label{sn6}
\end{equation}

\begin{equation}
S(u) = 0
\label{sn7}
\end{equation}

\begin{equation}
u_{a} (\vec{x},t) = u_{0}(\vec{x},t) + \sum\limits_{j=1}^{N} a_{j}(t) \phi _{j}(\vec{x})
\label{sn8}
\end{equation}

\noindent where $u_{a} (\vec{x},t)$ is the approximate solution, $\phi _{j}$ are known analytical functions, $a_{j}$ are coefficients to be determined, and $u_{0}(\vec{x},t)$ is chosen in order to satisfy the boundary and the initial conditions. By inserting Eq. \ref{sn8} in Eqs. \ref{sn6} and \ref{sn7}, and considering that $\phi _{j} = \phi _{j}(\vec{x},t)$, the coefficients $a _{j}$ become constants and Eq. \ref{sn6} is reduced to a system of algebraic equations, with a non-zero residual $R$ given by:

\begin{equation}
R(a_{0}, a_{1}, ... , a_{N},\vec{x}) = L(u_{a}) = L(u_{0}) + \sum\limits_{j=1}^{N} a_{j} L(\phi _{j})
\label{sn9}
\end{equation}

Once assumed that $u_{a}$ is a solutions of the problem,
	
\begin{equation}
R(a_{0}, a_{1}, ... , a_{N},\vec{x}) = L(u_{a}) = L(u_{0}) + \sum\limits_{j=1}^{N} a_{j} L(\phi _{j}) = 0
\label{sn10}
\end{equation}
	
\noindent and integrating Eq. \ref{sn10} in a proper domain $D$,
	
\begin{equation*}
\int_{D} R w_{k} d\vec{x} = \int_{D} \{ L(u_{0}) + \sum\limits_{j=1}^{N} a_{j} L(\phi _{j}) \} w_{k} d\vec{x} = 0 \Leftrightarrow
\end{equation*}

\begin{equation*}
\int_{D} R w_{k} d\vec{x} = \int_{D} L(u_{0})w_{k} d\vec{x} + \int_{D} \{\sum\limits_{j=1}^{N} a_{j} L(\phi _{j})w_{k}\} d\vec{x} = 0 \Leftrightarrow
\end{equation*}

\begin{equation*}
\int_{D} R w_{k} d\vec{x} = \int_{D} L(u_{0})w_{k} d\vec{x} + \sum\limits_{j=1}^{N} a_{j} \{ \int_{D} L(\phi _{j})w_{k} d\vec{x} \} = 0 \Leftrightarrow
\end{equation*}
	
\begin{equation}
<R,w_{k}> \ \ = \ \ <L(u_{0}),w_{k}> + \sum\limits_{j=1}^{N} a_{j} <L(\phi _{j}),w_{k}> \ \ = \ \ 0
\label{sn11}
\end{equation}

\noindent we obtain Eq. \ref{sn11}, which is true for any $w_{k}$. Therefore, the coefficients $a_{j}$ in Eq. \ref{sn8} are determined by solving the system of equations given by Eq. \ref{sn12} for $k = 0,1,...,N$,
	
\begin{equation}
<R,w_{k}> \ \ = \ \ 0
\label{sn12} 
\end{equation}
	
\noindent or,
	
\begin{equation}
\sum\limits_{j=1}^{N} a_{j} <L(\phi _{j}),w_{k}> = - <L(u_{0}),w_{k}>
\label{sn13}
\end{equation}
	
From Eq. \ref{sn13}, together with Eqs. \ref{sn6} to \ref{sn8}, we obtain the approximate solution $u_{a}$. The Galerkin method is a particular case of the Weighted Residual method, obtained when the trial functions $\phi _{k}$ are chosen from the same family of the base functions $w_{k}$ as in Eq. \ref{sn14},
	
\begin{equation}\label{sn14}
w_{k}(\vec{x}) = \phi _{k}(\vec{x})\ \ ,\ \ for \ \ k = 0,1,...,N.
\end{equation}

We developed a numerical approach based on a Galerkin method discretized with Chebyshev polynomials of the first kind, given by Eq. \ref{Chebyshev},

\begin{equation}
T_n (\cos \Gamma) = \cos (n\Gamma )
\label{Chebyshev}
\end{equation}

\noindent making it possible to express the Orr-Sommerfeld equation and their boundary conditions as a generalized eigenvalue problem. Because of the orthogonal properties of the Chebyshev polynomials in the interval $[-1;1]$, we transferred the problem domain to that interval by applying the transformation $z = 2y + 1$ for $y \in [-1;0]$. The choice of Chebyshev polynomials was made because of their high accuracy and their orthogonal properties, which makes the implementation easier.  Next, we rearranged the boundary conditions to eliminate $\hat{\eta}$, and discretized $\hat{\Psi}(z)$ by using Eq. \ref{sn43}
	
\begin{equation}
\hat{\Psi}(z) = \sum_{k=0}^{N} a_{k} T_{k}(z); \    \ k \in \{\mathbb{Z} \ \ | \ \ k \geq 0 \}
\label{sn43}
\end{equation}

\noindent In addition, we applied the inner products of functions given by Eq. $\ref{sn43}$. With this procedure, the Orr-Sommerfeld equation is written in terms of the Chebyshev polynomials. For the boundary conditions, we simply applied Eq. $\ref{sn43}$ to the wall and interface, at $z=-1$ and $z=+1$, respectively. At the end, we obtained the eigenvalue problem given by Eq. \ref{eigenproblem},

\begin{equation}
[\textbf{A}]_{NxN} \vec{a} = c [\textbf{B}]_{NxN} \vec{a}
\label{eigenproblem}
\end{equation}
	
\noindent where $N$ is the number of Chebyshev polynomials to be used, and the matrices $\textbf{A}$ and $\textbf{B}$ are written as $\textbf{A} = A_{r} + i A_{i}$ and $\textbf{B} = B_{r} + i B_{i}$, respectively.  One of the advantages of this method is its readiness in including the boundary conditions in the discretized equations, which is done by replacing the last lines of Eq. \ref{eigenproblem} with the transformed boundary conditions. 

A numerical code was written in Matlab environment to solve Eq. \ref{eigenproblem}. We used the function \textit{eig}, which uses a Cholesky factorization or a QZ algorithm (generalized Schur decomposition) based on the properties of $\textbf{A}$ and $\textbf{B}$. If $\textbf{A}$ and $\textbf{B}$ are symmetric, the standard choice will be the Cholesky factorization, otherwise the function will use a QZ algorithm. In order to optimize our code, we implemented an Inverse Iteration method \cite{hossain2011convection}, that tracks the physical eigenvalue based on an initial guess. The Inverse Iteration method is useful to produce the neutral stability curves, once that only the physical eigenvalues and the respective eigenvectors are tracked on each iteration. Therefore, the numerical code finds the entire eigenvalue spectrum from an initial guess. For the results presented in this work we used $N$ = 80 unless otherwise specified.

\section{Results}
\label{section:results} 

In order to evaluate our numerical results, we developed an asymptotic expansion for long waves as done by \cite{benney1966long}, and obtained similar results. For a long wave disturbance, the wavenumber $\alpha$ can be treated as a small parameter; therefore, we expanded the eigenfunction $\hat{\Psi}(y)$ and the eigenvalue $c$ in power series of $\alpha$, from $O(1)$ to $O(\alpha^2)$, where $O$ stands for \textit{order}, as shown in Eqs \ref{as2} and \ref{as3}. 

\begin{equation}
\hat{\Psi}(y) = \hat{\Psi}_{0}(y) + \alpha\hat{\Psi}_{1}(y) + \alpha^2\hat{\Psi}_{2}(y) + O(\alpha^3)
\label{as2}
\end{equation}
	
\begin{equation}
c = c_{0} + \alpha c_{1} + \alpha^2 c_{2} + O(\alpha^3)
\label{as3}
\end{equation}

At $O(1)$, we found that the eigenvalue $c_ {0}$ is real and independent of the wavenumber. Since the imaginary part of $c_ {0}$ is zero, the growth rate is zero; therefore, the initial perturbation has a non-zero celerity, but it does not grow nor decreases at $O(1)$. At $O(\alpha)$, the solution is $c_{1} = iRe\frac{8}{15}[1 - \frac{5}{8}(\frac{1}{Fr} + \frac{\alpha^2}{We})]$, which is purely imaginary, affecting the growth rate but not the celerity of initial perturbations. The most unstable wavenumber is obtained at this order. This expression agrees with Benney's results for $O(\alpha)$, which is given by $c_{1} = iRe(Re - \frac{5}{4}cot(\theta)) = iRe ^{2}(1 - \frac{5}{8}\frac{1}{Fr})$. Benney neglected the contribution of Weber number until $O(\alpha ^3)$; however, the same criteria for the onset of instabilities was obtained. The development of these equations and the results for $O(\alpha^2)$ are presented in Appendix \ref{appendix_A}. With $c_{1}$, it is possible to write the growth rate $\sigma = \alpha c_{i} = \alpha^2 c_{1i}$ in the form,

\begin{equation}
	\sigma = \frac{\alpha^2 Re}{3}\Biggl(\frac{1}{Fr_{c}} - \frac{1}{Fr}\Biggl) - \frac{Re}{3We}\alpha^4
	\label{as60}
\end{equation}
	
\noindent where $Fr_{c} = \frac{5}{8}$. When $Fr < Fr_{c}$, $\sigma < 0$ for every $\alpha$, and the liquid film is stable. For $Fr > Fr_{c}$, perturbations with $\alpha < \alpha_{c}$ are amplified, and the liquid film is linearly unstable. Perturbations with $\alpha > \alpha_{c}$ are attenuated by the combined effects of surface tension and viscosity. By considering $\sigma = 0$, we obtain $\alpha_{c}$, given by Eq. \ref{as62},
	
\begin{equation}
	\alpha_{c}^2 = We\Biggl(\frac{1}{Fr_{c}} - \frac{1}{Fr}\Biggl)
	\label{as62}
\end{equation}
	
	\noindent where $Fr_{c} = \frac{5}{8}$. The critical Froude number, $Fr_{c} = \frac{5}{8}$, is the value above which the liquid film is linearly unstable. Finally, we can obtain the unstable band for large Froude numbers, 
	
\begin{equation}
\lim_{Fr\to\infty} \alpha_{c} = \lim_{Fr\to\infty} \sqrt{We \Biggl(\frac{1}{Fr_{c}} - \frac{1}{Fr} \Biggl)} \rightarrow \sqrt{\frac{We}{Fr_{c}}}
\label{as63}
\end{equation}
	
\noindent According to Eq. \ref{as63}, the unstable band is limited for large Froude numbers.

Next, we present the numerical results obtained with our method and compare them with analytical results obtained with Perturbation Methods as well as with some other numerical results. In order to test the convergence of the numerical method, we used as reference a case presented by Charru \cite{Livre_Charru}, for which $\theta = \pi / 3$, $\alpha$ = 0.01, $We$ = 0.0001 and $Re$ = 1. Tables $\ref{tab1}$ and $\ref{tab2}$ show the results for the physical eigenvalue $c = c_{r} + ic_{i}$ obtained by varying the number of Chebyshev polynomials, $N$, without and with the Inverse Iteration method \cite{hossain2011convection}, respectively. Without the Inverse Iteration method, we obtained convergence of the results with 10 polynomials to $c_r = 1.999818$ and $\sigma / \alpha^2 = - 0.185036$, which is the same result obtained by Charru \cite{Livre_Charru} using a Spectral Collocation method with 16 Chebyshev polynomials. If the number of polynomials is increased above 15, the numerical error in the solution increases. In order to overcome this problem and find a better result, we implemented an Inverse Iteration method. With the Inverse Iteration method, the numerical error does not increase for $N>15$, and full convergence up to 16 decimals is achieved for $c_i$ with $N \geq 20$. The Galerkin method with Inverse Iteration solves the system of equations and then uses the result as a first approximation of the linear system for the next computation. The choice of using $N$ = 80 in most of the results presented in this work was arbitrary, once it is only necessary 20 polynomials to obtain good results.

\begin{table}[!ht]	
	\begin{center}
		\begin{tabular}{|c|c|c|c|}
			\hline
			N & $c_{r}$ & $c_{i}$ \\\hline 
			$4$ & 1.999756099032411 & -0.000518494614339 \\\hline
			$6$ & 1.999817050142938 & -0.001850362515255 \\\hline
			$8$ & 1.999817637205630 & -0.001850361243183 \\\hline
			$10$ & 1.999817653549203 & -0.001850361091185 \\\hline
			$15$ & 1.999817653549203 & -0.001850361091185 \\\hline
			$20$ & 1.999817653872101 & -0.001850360665793 \\\hline
			$30$ & 1.999817656051497 & -0.001850368834596 \\\hline
			$40$ & 1.999818070722873 & -0.001847606451498 \\\hline
			$60$ & 1.999817868510843 & -0.001853477042879 \\\hline
			$80$ & 1.999763700422891 & -0.001812682637214 \\\hline
		\end{tabular}
	\end{center}
	\caption{Numerical results for the physical eigenvalue as function of the number of Chebyshev polynomials. All the results were obtained without the Inverse Iteration method.}
	\label{tab1}
\end{table}

\begin{table}[!ht]	
	\begin{center}
		\begin{tabular}{|c|c|c|c|}
			\hline
			N & $c_{r}$ & $c_{i}$ \\\hline 
			$4$ & 1.999756099032410 & -0.000518494614338 \\\hline
			$6$ & 1.999817050142902 & -0.001850362515039 \\\hline
			$8$ & 1.999817637204247 & -0.001850361243036 \\\hline
			$10$ & 1.999817653549002 & -0.001850361090049 \\\hline
			$15$ & 1.999817653549107 & -0.001850361088397 \\\hline
			$20$ & 1.999817653548911 & -0.001850361090320 \\\hline
			$30$ & 1.999817653548911 & -0.001850361090320 \\\hline
			$40$ & 1.999817653548911 & -0.001850361090320 \\\hline
			$60$ & 1.999817653548912 & -0.001850361090320 \\\hline
			$80$ & 1.999817653548912 & -0.001850361090320 \\\hline
		\end{tabular}
	\end{center}
	\caption{Numerical results for the physical eigenvalue as function of the number of Chebyshev polynomials. All the results were obtained with the Inverse Iteration method \cite{hossain2011convection}.}	
	\label{tab2}	
\end{table}

\begin{figure}[!ht]
\centering
\includegraphics[width=0.70\textwidth]{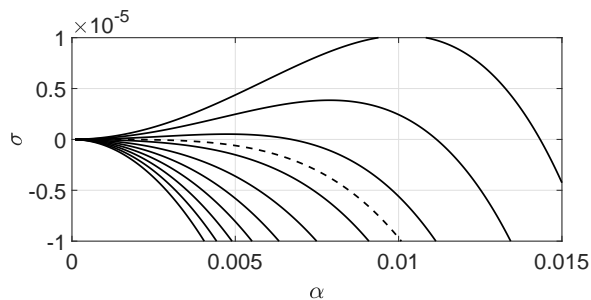}
\caption{Dispersion relation $\sigma (\alpha)$ parametrized by $\theta$, plotted with the numerical data. The dashed line represents the growth rate for $\theta_{c}$, equal to 28.36$^\circ$ in this case.}
\label{Disp_relation}
\end{figure}

Figure \ref{Disp_relation} shows the numerical results for the dispersion relation of initial instabilities, $\sigma (\alpha)$, parametrized by $\theta$. The results for both the analytical and numerical solutions were computed with the reference values $\mu$ = 0.001 N s/m$^{2}$, $\rho$ = 998.2071 Kg/m$^{3}$, $g$ = 10 m/s$^{2}$, $\gamma$ = 0.07275 N/m and 0.1 mm thickness. In Fig. \ref{Disp_relation}, we used $\frac{\pi}{8} < \theta < \frac{\pi}{5.8}$ in order to consider  $\theta < \theta_{c}$ and $\theta > \theta_{c}$. The numerical and the analytical results are in perfect agreement and the results are perfectly superposed; therefore, Fig. \ref{Disp_relation} shows only the numerical results.

\begin{figure}[!ht]
\centering
\includegraphics[width=0.9\textwidth]{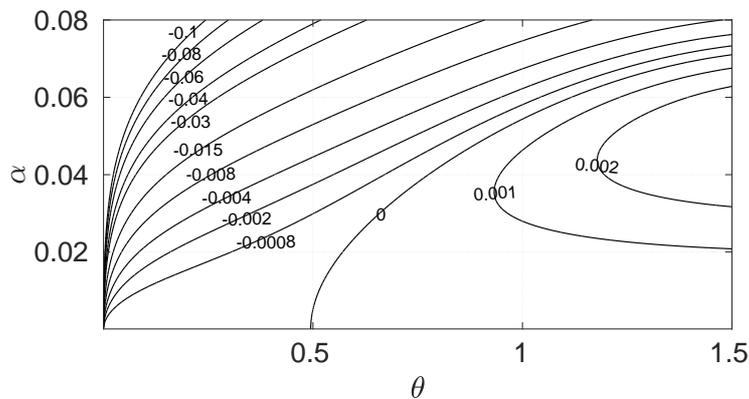}
\caption{Stability diagram plotted with the numerical results for $h$ = 0.01 mm.}
\label{Stabil_diagram}
\end{figure}

Figure \ref{Stabil_diagram} presents the stability diagram plotted with the numerical results for $h$ = 0.01 mm and $0 < \theta < 1.5$. Each plotted curve corresponds to a specific growth rate $\sigma$, and the curve for $\sigma = 0$ is the marginal stability curve, which separates the stable and unstable domains. These domains have negative and positive values of the growth rate, respectively. The diagram shows that the width of the unstable band is smaller at the threshold $\theta = \theta_{c}$, and that the growth rate presents different behaviors according to $\alpha$. We note here that this is a numerical solution of the linear stability equations and, therefore, the solutions are valid at the onset of instability.

\begin{figure}[!ht]
\centering
\includegraphics[width=0.8\textwidth]{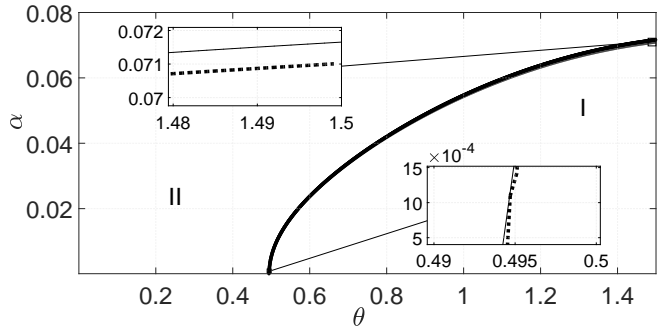}
\caption{Marginal stability curve for asymptotic and numerical solutions. The continuous and dotted lines correspond to the asymptotic and numerical solutions, respectively.}
\label{marginal_curve}
\end{figure}

Figure \ref{marginal_curve} presents a comparison between the asymptotic and numerical solutions for the marginal stability curve, $\sigma = 0$, for $0 < \theta < 1.5$. This curve separates the unstable and stable domains, represented by $I$ and $II$, respectively. The continuous and dotted lines correspond to the asymptotic and numerical solutions, respectively. The agreement is good, especially for $\theta < 1$. For $\theta > 1$, the solutions diverge slightly. However, Eq. \ref{as63} predicts a limit of $\alpha _{c} = 0.0705$ for the wavenumber, and the solutions are in good agreement with respect to this limit.

\begin{figure}[!ht]
\centering
\includegraphics[width=0.80\textwidth]{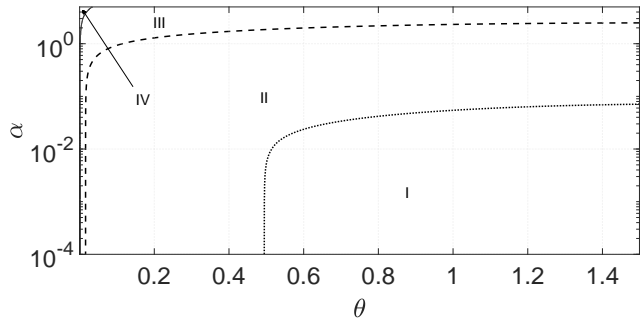}
\caption{Marginal stability curves for three different values of the thickness $h$. The continuous, dashed and dotted lines correspond to $h$ equal $10$ mm, $1$ mm and $0.1$ mm, respectively.}
\label{marginal_curves_diff_thick}
\end{figure}

Figure \ref{marginal_curves_diff_thick} presents the marginal stability curves for three different values of the film thickness, showing four different stability regions. The dotted curve corresponds to a liquid layer with thickness equal $h$ = 0.1 mm, and $I$ represents the unstable region while $II$, $III$ and $IV$ represent the stable region. The dashed curve corresponds to $h$ = 1 mm, with $I$ and $II$ being the unstable region and $III$ and $IV$ the stable one. The continuous curve corresponds to $h$ = 10 mm, with $I$, $II$ and $III$ being the unstable region and $IV$ the stable one. These three curves show that when the liquid layer is thicker, the unstable band is larger and begins at smaller values of the slope angle. The diagram shown in Fig. \ref{marginal_curves_diff_thick} is a good representation of how the film thickness affects the growth rate of the instabilities. For higher values of $h$, the effect of the inertia is more pronounced and the angle necessary to reach the critical conditions is lower.

We note that the diagram shown in Fig. \ref{marginal_curves_diff_thick} was plotted with the use of the Inverse Iteration method \cite{hossain2011convection} in association with the Galerkin Method. Figure \ref{marginal_curves_diff_thick} is very useful because it shows the unstable bands in function of the inclination angle for different film thicknesses, and therefore, for a given liquid, the formation of Kapitza waves is accessed directly from the physical parameters $h$ and $\theta$. From the authors' knowledge, this is the first time that this kind of diagram is presented, giving new possibilities for a direct analysis of the problem.

\begin{figure}[!ht]
	\centering
	\includegraphics[width=0.70\textwidth]{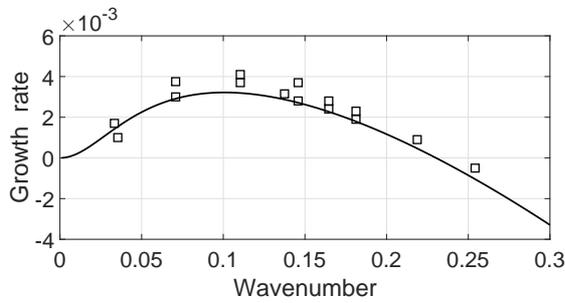}
	\caption{Comparison between the numerical solution of the dimensionless spatial growth rate and the experimental data produced by Liu et al. \cite{liu1993measurements}. The solid line corresponds to the numerical result and the symbols to the experimental data.}
	\label{num_exp}
\end{figure}

Finally, we compare our numerical results with some experimental results. Figure \ref{num_exp} presents our numerical results and the experimental results of Liu et al. \cite{liu1993measurements} for the spatial growth rate of Kapitza waves. The continuous line corresponds to the numerical results and the symbols to the experimental results. The numerical results were obtained with the same parameters of the experiments, which are $\rho$ = 1.13 g/cm$^{3}$, $g$ = 9.80665 m/s$^{2}$ and $\theta$ = 4.6$^{o}$. For the thickness of the liquid film, we used Liu's Weber number equal to $63$, defined differently from ours, and we found $h$ = 1.119 cm. Using these parameters, together with $\nu$ = 4.89 $\cdot$ 10 $^{-6}$ m$^{2}$/s and $\gamma$ = 69 $\cdot$ 10$^{-3}$N/m, we found $We = 0.19$ from our definition of the Weber number, and $Re = 23$. We obtained the data for the temporal growth rate as a function of the wavenumber with our numerical code. In order to compare our results with Liu's data, we applied to our data the Gaster relation \cite{Gaster} $\omega _{i} ^{T} = -c_{g}k_{i} ^{S}$, which relates the temporal growth rate $\omega _{i} ^{T}$ to the spatial growth rate $k_{i} ^{S}$ through the group velocity $c_{g} = - \frac{\partial \omega _{r}}{\partial k _{r}}$. From the asymptotic solution at $O(1)$, we found the dimensionless group velocity $c_{g} = 2$ and the dimensionless Gaster relation becomes $\omega _{i} ^{T} = 2k_{i} ^{S}$. Using this last equation, we found the numerical solution for the spatial growth rate.

From Fig. \ref{num_exp}, we observe that the agreement of our solution with the experimental data of Liu et al. \cite{liu1993measurements} is very good.

\section{Conclusions}
\label{section:conclusions}

This paper presented a spectral method to solve the Orr-Sommerfeld equation with free surface boundary conditions in order to numerically address the formation of Kapitza waves on liquid films. Our numerical approach is based on a Galerkin method with Chebyshev polynomials of the first kind, making it possible to express the Orr-Sommerfeld equation and their boundary conditions as a generalized eigenvalue problem. We combined the Galerkin method with the Inverse Iteration method \cite{hossain2011convection}, which tracks the physical eigenvalue based on an initial guess. The present method is at the same time straightforward in including the boundary conditions in the discretized equations and stable. We compared our numerical results with analytical solutions based on Perturbation Methods, which are valid only for long wave instabilities, and showed that the results agree in the region of validity of the long-wave hypothesis. A comparison with the Spectral Collocation method showed that the present method converges with a smaller number of Chebyshev polynomials. Also, our method reaches full convergence up to 16 decimals for the growth rate with $N \geq 20$. In addition, we compared our results with the experimental results of Liu et al. \cite{liu1993measurements}, and the agreement between them is very good. With the present method, different from previous ones, it is possible to access the formation of Kapitza waves directly from the physical parameters $h$ and $\theta$ for all physical eigenvalues. This gives new possibilities for a direct analysis of the problem. The method is stable, fast, and capable to solve initial instabilities in free surface flows for all ranges of $h$ and $\theta$ by just setting the fluid properties.



\begin{acknowledgements}
Bruno Chimetta is grateful to the Emerging Leaders in the Americas Program (ELAP) and to Capes for the scholarship grants. Mohammad Hossain is grateful to Western University for providing some computational resources. Erick Franklin is grateful to FAPESP (grant no. 2016/13474-9), to CNPq (grant no. 400284/2016-2) and to FAEPEX/UNICAMP (conv. 519.292) for the provided financial support.
\end{acknowledgements}


\bibliography{references}
\bibliographystyle{spphys}

\appendix
\section{Asymptotic solution}
\label{appendix_A}

This appendix is devoted to asymptotic solutions of the equation of Orr-Sommerfeld together with the boundary conditions of the problem. To find these solutions we expand the eigenfunction $\hat{\Psi}(y)$ and the eigenvalue $c$ in power series of $\alpha$, from $O(1)$ to $O(\alpha^2)$.
	
\subsection{Solution for $O(1)$}

Considering,
	
\begin{equation}
\alpha << 1 ; \  \ Re = \textit{O}(1) ; \   \ \frac{We}{\alpha^2} = \textit{O}(1)
\label{appendix1}
\end{equation}

For a long wave disturbance, the wavenumber $\alpha$ can be treated as a small parameter. Equations \ref{Orr-Sommerfeld} to \ref{cc5} suggest that the speed $c$ and the amplitude $\hat{\Psi}$ of eigenfunctions can be treated as a power series of $\alpha$ as follows:

\begin{equation}
\hat{\Psi}(y) = \hat{\Psi}_{0}(y) + \alpha\hat{\Psi}_{1}(y) + \alpha^2\hat{\Psi}_{2}(y) + ...
\label{appendix2}
\end{equation}

\begin{equation}
c = c_{0} + \alpha c_{1} + \alpha^2 c_{2} + ...
\label{appendix3}
\end{equation}
	
In order to find an approximate solution, we replace Eqs. \ref{appendix2} and \ref{appendix3} into the Orr-Sommerfeld equation and the terms of the same order are collected. The same procedure is applied on the boundary conditions.

At $O(1)$:

\begin{equation}
D^4\hat{\Psi}_{0}(y) = 0
\label{appendix4}
\end{equation}
	
\begin{equation}\label{appendix14}
\hat{\Psi}_{0}(-1) = 0
\end{equation}

\begin{equation}\label{appendix15}
D\hat{\Psi}_{0}(-1) = 0
\end{equation}

\begin{equation}\label{appendix7}
\hat{\Psi}_{0}(0) - (c_{0} - 1)\hat{\eta} = 0
\end{equation}

\begin{equation}\label{appendix16}
D^3\hat{\Psi}_{0}(0) = 0
\end{equation}

\begin{equation}\label{appendix17}
D^2\hat{\Psi}_{0}(0) - 2\frac{\hat{\Psi}_{0}(0)}{c_{0} - 1} = 0
\end{equation}

From Eqs. \ref{appendix4} to \ref{appendix17},

\begin{equation}
\hat{\Psi}_{0}(y) = \hat{\eta}(y + 1)^2 ;\   \ c_{0} = 2
\label{appendix28}
\end{equation}

\noindent where $\hat{\eta}$ is the amplitude of the interface deformation. The eigenvalue $c_ {0}$ is real and independent of wavenumber; therefore, all disturbances are propagated with the same speed $2U_{0}$, independent of the wavelength (non-dispersive). Since the imaginary part of $c_ {0}$ is zero, the growth rate of instability is zero, and there is no instability at $O(1)$.

\subsection{Solution for $O(\alpha)$}

At $O(\alpha)$:

\begin{equation}
D^4\hat{\Psi}_{1}(y) = 4iRe\hat{\eta}y
\label{appendix46}
\end{equation}

\begin{equation}
\hat{\Psi}_{1}(-1) = 0
\label{appendix41}
\end{equation}

\begin{equation}
D\hat{\Psi}_{1}(-1) = 0
\label{appendix42}
\end{equation}

\begin{equation}
\hat{\Psi}_{1}(0) = c_{1}\hat{\eta}
\label{appendix35}
\end{equation}

\begin{equation}
D^2\hat{\Psi}_{1}(0) = 0
\label{appendix43}
\end{equation}

\begin{equation}
D^3\hat{\Psi}_{1}(0) = -2iRe \hat{\eta} + iRe \hat{\eta}\Biggl(\frac{1}{Fr} + \frac{\alpha^2}{We}\Biggl)
\label{appendix44}
\end{equation}

By integrating Eq. $\ref{appendix46}$, and applying Eq. \ref{appendix41} to Eq. \ref{appendix44} we find:

\begin{equation}
\begin{split}
\hat{\Psi}_{1}(y) = & iRe\hat{\eta}\Biggl \lbrace \frac{y^5}{30} + \Biggl[-\frac{1}{3} + \frac{1}{6}\Biggl(\frac{1}{Fr} + \frac{\alpha^2}{We}\Biggl)\Biggl]y^3 + \Biggl[\frac{5}{6} - \frac{1}{2}\Biggl(\frac{1}{Fr} + \frac{\alpha^2}{We}\Biggl)\Biggl]y + \\
& + \frac{8}{15}\Biggl[1 - \frac{5}{8}\Biggl(\frac{1}{Fr} + \frac{\alpha^2}{We}\Biggl)\Biggl]\Biggl \rbrace
\end{split}
\label{appendix56}
\end{equation}
	
\begin{equation}
c_{1} = iRe\frac{8}{15}\Biggl[1 - \frac{5}{8}\Biggl(\frac{1}{Fr} + \frac{\alpha^2}{We}\Biggl)\Biggl]
\label{appendix57}
\end{equation}

At $O(1)$ the solution  is purely imaginary, and does not contribute to the wave speed, but affects the growth rate $\sigma = \alpha c_{i} = \alpha^2 c_{1i}$ significantly,

\begin{equation}
\begin{split}
\sigma = \alpha^2 c_{1i} = \alpha^2 Re\frac{8}{15}\Biggl[1 - \frac{5}{8}\Biggl(\frac{1}{Fr} + \frac{\alpha^2}{We}\Biggl)\Biggl] = \\
= \frac{\alpha^2 Re}{3}\Biggl(\frac{1}{\frac{5}{8}} - \frac{1}{Fr}\Biggl) - \frac{Re}{3We}\alpha^4
\end{split}
\label{appendix59}
\end{equation}

Therefore, we can write,
	
\begin{equation}
\sigma = \frac{\alpha^2 Re}{3}\Biggl(\frac{1}{Fr_{c}} - \frac{1}{Fr}\Biggl) - \frac{Re}{3We}\alpha^4
\label{appendix60}
\end{equation}

\noindent where,

\begin{equation}
Fr_{c} = \frac{5}{8}
\end{equation}
	
When $Fr < Fr_{c}$, $\sigma$ is negative for every $\alpha$, and the flow of the liquid film is stable. For $Fr > Fr_{c}$, perturbations of wavenumber below $\alpha_{c}$ will be amplified. We can find $\alpha_{c}$ by,
	
\begin{equation}
\sigma = 0 \Leftrightarrow \alpha^2\Biggl[\frac{Re}{3}\Biggl(\frac{1}{Fr_{c}} - \frac{1}{Fr}\Biggl) - \frac{Re}{3We}\alpha^2 \Biggl] = 0
\label{appendix61}
\end{equation}
	
Excluding the case $\alpha^2 = 0$ we obtain,
	
\begin{equation}
\frac{Re}{3}\Biggl(\frac{1}{Fr_{c}} - \frac{1}{Fr}\Biggl) - \frac{Re}{3We}\alpha^2 = 0 \Leftrightarrow \alpha_{c}^2 = We\Biggl(\frac{1}{Fr_{c}} - \frac{1}{Fr}\Biggl)
\label{appendix62}
\end{equation}
	
Perturbations with wavenumber $\alpha > \alpha_{c}$ are attenuated due to the combined effect of surface tension and viscosity. The number $Fr_{c} = \frac{5}{8}$ is the critical Froude number above which the liquid film is linearly unstable.

\subsection{Solution for $O(\alpha^{2})$}

At $O(\alpha^{2})$:

\begin{equation}
\begin{split}
D^4\hat{\Psi}_{2}(y) = 4\hat{\eta} - Re^2\hat{\eta}\Biggl \lbrace -\frac{3}{5}y^5 + \Biggl[\frac{2}{3} - \frac{2}{3}\Biggl(\frac{1}{Fr} + \frac{\alpha^2}{We}\Biggl)\Biggl]y^3 + \\
+ \Biggl[\frac{11}{3} - 2\Biggl(\frac{1}{Fr} + \frac{\alpha^2}{We}\Biggl)\Biggl]y\Biggl \rbrace
\end{split}
\label{appendix67}
\end{equation}

\begin{equation}
\hat{\Psi}_{2}(-1) = 0
\label{appendix71}
\end{equation}

\begin{equation}
D\hat{\Psi}_{2}(-1) = 0
\label{appendix72}
\end{equation}

\begin{equation}
\hat{\Psi}_{2}(0) = \hat{\eta}c_{2}
\label{appendix74}
\end{equation}

\begin{equation}
D^2\hat{\Psi}_{2}(0) = - \hat{\eta}
\label{appendix75}
\end{equation}

\begin{equation}
D^3\hat{\Psi}_{2}(0) = 6\hat{\eta} + Re^2\hat{\eta}\Biggl[\frac{121}{30} - \frac{5}{2}\Biggl(\frac{1}{Fr} + \frac{\alpha^2}{We}\Biggl)\Biggl]
\label{appendix77}
\end{equation}

Integrating Eq. \ref{appendix67} and replacing Eq. \ref{appendix71} - Eq. \ref{appendix77} we obtain,

\begin{equation}
\begin{split}
\hat{\Psi}_{2}(y) = \frac{\hat{\eta}}{6}y^4 - \frac{Re^2\hat{\eta}}{60}\Biggl \lbrace-\frac{1}{84}y^9 + \frac{1}{21}\Biggl[1 - \Biggl(\frac{1}{Fr} + \frac{\alpha^2}{We}\Biggl)\Biggl]y^7 + \\
+ \Biggl[\frac{11}{6} - \Biggl(\frac{1}{Fr} + \frac{\alpha^2}{We}\Biggl)\Biggl]y^5\Biggl \rbrace + \frac{A}{6}y^3 + \frac{B}{2}y^2 + Cy + D
\end{split}
\label{appendix78}
\end{equation}

\begin{equation}
c_{2} = - 2 - \frac{32}{63}Re^2\Biggl[1 - \frac{5}{8}\Biggl(\frac{1}{Fr} + \frac{\alpha^2}{We}\Biggl)\Biggl]
\label{appendix84}
\end{equation}

\noindent where A, B, C and D are, respectively,
	
\begin{equation}
A = 6\hat{\eta} + Re^2\hat{\eta}\Biggl[\frac{57}{30} - \frac{7}{6}\Biggl(\frac{1}{Fr} + \frac{\alpha^2}{We}\Biggl)\Biggl]
\label{appendix82}	
\end{equation}

\begin{equation}
B = -\hat{\eta}
\label{appendix81}
\end{equation}

\begin{equation}
C = -\frac{10\hat{\eta}}{3} - \frac{Re^2\hat{\eta}}{60}\Biggl[\frac{1333}{28} - \frac{89}{3}\Biggl(\frac{1}{Fr} + \frac{\alpha^2}{We}\Biggl)\Biggl]
\label{appendix83}
\end{equation}

\begin{equation}
D = \hat{\eta}c_{2}
\label{appendix80}
\end{equation}
	
Performing the calculations at $O(\alpha ^{2})$ we find a correction for the real part of eigenvalue $c$. This correction only affects the wave speed, therefore, at $O(\alpha ^{2})$, long wavelengths are weakly dispersive \cite{benney1966long}.

\end{document}